\documentclass[%
 reprint,
 amsmath,amssymb,
aps,
prl,
]{revtex4-1}

\usepackage{graphicx}% Include figure files
\usepackage{dcolumn}% Align table columns on decimal point
\usepackage{bm}% bold math
\usepackage{amsmath}
\newcommand{\EP}{{\it{e}\rm{-ph}}}
\newcommand{\EE}{{\it{e-e}}}

\begin{document}

%\preprint{APS/123-QED}

\title{The electronic thermal conductivity of graphene}% Force line breaks with \\

\author{Tae Yun Kim}
\affiliation{Department of Physics, Seoul National University, Seoul 08826, Korea}

\author{Cheol-Hwan Park}
\email{cheolhwan@snu.ac.kr}
\affiliation{Department of Physics, Seoul National University, Seoul 08826, Korea}

\author{Nicola Marzari}
\affiliation{Theory and Simulation of Materials (THEOS), and National Centre for Computational Design and Discovery of Novel Materials (MARVEL), \'Ecole Polytechnique F\'ed\'erale de Lausanne, 1015 Lausanne, Switzerland}

\date{\today}% It is always \today, today,
             %  but any date may be explicitly specified

\begin{abstract}
Graphene, as a semimetal with the largest known thermal conductivity,
is an ideal system to study the interplay between electronic and lattice contributions
to thermal transport. While the total electrical and thermal conductivity have
been extensively investigated, a detailed first-principles study of its electronic thermal 
conductivity
is still missing. Here, we first characterize the 
electron-phonon
intrinsic contribution 
to the electronic thermal resistivity of graphene as a function of doping using
electronic and phonon dispersions and electron-phonon couplings
calculated from first principles at the level of density-functional theory
and many-body perturbation theory ({\it GW}). 
Then, we include
extrinsic electron-impurity scattering using low-temperature experimental estimates.
Under these conditions, we find that the in-plane electronic thermal conductivity $\kappa_e$
of doped graphene is $\sim$300~W/mK at room temperature, independently of
doping. This result is much larger than expected, and comparable to 
the total thermal conductivity of typical metals, contributing  
$\sim$10~\% to the total thermal conductivity of bulk graphene.
Notably, in samples whose physical or domain sizes are of the order of few 
micrometers or smaller, the relative contribution coming from the electronic 
thermal conductivity is more important
than in the bulk limit, since
lattice thermal conductivity is much more sensitive to sample or grain size at 
these scales. Last, when electron-impurity scattering effects are included, we find that
the electronic thermal conductivity is reduced by 30 to 70~\%. We also find that
the Wiedemann-Franz law is broadly satisfied at low and high temperatures, but 
with the largest deviations of 20--50~\% around room temperature.
\end{abstract}

%Use showkeys class option if keyword
                              %display desired
\maketitle

%\tableofcontents

%%%MAIN TEXT%%%%
The thermal conductivity of graphene is extremely high, which is not only fascinating
from the scientific point of view, but is also promising for many technological 
applications. So far, a fairly wide range of thermal conductivities have been reported experimentally~\cite{balandin2008, chen_raman_2011, seol2010, cai2010}, with the measured thermal conductivity of suspended graphene at room temperature
ranging from 2600 to 5300~W/mK~\cite{balandin2008, chen_raman_2011}, which is higher than that of any other known material.
The measured thermal conductivity of graphene supported on a substrate is much lower (370--600~W/mK) than that of the suspended case, but still comparable to or higher than that of typical metals~\cite{seol2010, cai2010}.
It is widely assumed that most of the thermal conduction is carried
by phonons~\cite{nika2009} and that the electronic contribution is negligible, with
experiments hinting that the electronic thermal conductivity $\kappa_e$
obtained from the measured electrical conductivity by applying the Wiedemann-Franz law
could be as low as 1~\% of the total thermal conductivity~\cite{ghosh2008}.

In typical metals, the total thermal conductivity is the sum of the electronic contribution $\kappa_e$ and the phonon contribution $\kappa_{\rm ph}$.
The kinetic theory of electrons and phonons provides a qualitative description of the temperature dependence of $\kappa_e$ and $\kappa_{\rm ph}$
in the low- and high-temperature limits. The electronic thermal conductivity is given by 
$\kappa_e = \frac{1}{3}\, C_e\, v_{\rm F}\, \Lambda_e$,
where $C_e$ is the specific heat of the electrons, $v_{\rm F}$ is the
electron group velocity, and $\Lambda_e$ the mean free path; similarly for phonons the
thermal conductivity is given by 
$\kappa_{\rm ph} = \frac{1}{3}\, C_{\rm ph}\,v_{\rm ph}\, \Lambda_{\rm ph}$,
where $C_{\rm ph}$ is the specific heat of phonons, $v_{\rm ph}$ the
phonon group velocity, and $\Lambda_{\rm ph}$ their mean free path.

In the low-temperature limit, electrons are 
scattered dominantly by impurities, which makes $\Lambda_e$ temperature-independent
and $\kappa_e\propto T$.
As the temperature increases, the number of phonons increases and electron-phonon (\EP) scattering limits $\kappa_e$.
The number of phonons is proportional to $T$
if $T$ is higher than the Debye temperature, $\Theta_{\rm D}$;
in this high-temperature limit, the mean free path of electrons $\Lambda_e$,
which is inversely proportional to the number of phonons, is
proportional to $1/T$, and $\kappa_e$ is temperature-independent. 
At low temperatures, phonons are scattered mostly by impurities,
defects, boundaries, etc., and $\Lambda_{\rm ph}$ is temperature-independent.
According to the Debye model in two and three dimensions,
$\kappa_{\rm ph} \propto C_{\rm ph} \propto T^{2}$ and $\kappa_{\rm ph} \propto C_{\rm ph}\propto T^{3}$ at low temperatures
$(T < \Theta_{\rm D})$, respectively.
(The corresponding relations for the quadratic flexural phonon branch in two
dimension~\cite{PhysRevB.76.115409} are $C_{\rm ph} \propto T$
and $\kappa_{\rm ph} \propto T^{1.5}$.)
At high temperatures $(T > \Theta_{\rm D})$, the specific heat of a phonon gas
is constant (Dulong-Petit law).
Since the dominant scattering mechanism in this limit is
the phonon-phonon (anharmonic) interaction, $\Lambda_{\rm ph}\propto 1/T$
when $T > \Theta_{\rm{D}}$ and $\kappa_{\rm ph}\propto 1/T$.
Therefore, in both the low- and high-temperature limits, the electronic contribution to the thermal conductivity is important.

For the electrical resistivity $1/\sigma$ doped graphene can be viewed as a 
two-dimensional metal, with a crossover as a function of temperature:
at low temperatures $1/\sigma \propto T^4$ while at high temperatures
$1/\sigma \propto T$. The crossover takes place at
the Bloch-Gr\"uneisen temperature
$T_{\rm{BG}} = 2\hbar{}v_{\rm{ph}}k_{\rm{F}}/k_{\rm{B}}$,
where $k_{\rm F}$ is the Fermi wavevector and $k_{\rm B}$ the Boltzmann constant.
The physics of the Bloch-Gr\"uneisen crossover has been studied
theoretically~\cite{hwang2008} and experimentally~\cite{efetov2010} in great detail, 
and we presented detailed first-principles studies of the \EP\ interactions and the 
intrinsic electrical resistivity of graphene, incorporating the effects of both 
low- and high-energy phonons~\cite{park2014NL,Sohier2014PRB}, explaining quantitatively
the experimental results~\cite{efetov2010}.

There has been a
theoretical study of the electrical thermal conductivity
$\kappa_e$~\cite{munoz2012}, based on the effects of impurities and \EP\ 
interactions described by an effective deformation potential, 
considering only longitudinal acoustic phonons.
However, the full consideration of the contributions on $\kappa_e$
from all the phonon branches and the detailed dependence
of the \EP\ interactions on the electronic and phonon wavevectors and band indices
at the first-principles level is still missing, and is presented here.
In particular, we employ an accurate first-principles fit (Ref.~\cite{park2014NL}) of 
the \EP\ coupling matrix elements for both the low-energy acoustic phonons
and the high-energy phonons of graphene that captures all the details of
the first-principles calculations, including many-body effects for
the electron-electron (\EE) interactions at the level of many-body perturbation theory
(the \textit{GW} approximation).
Our paper presents the first comprehensive calculation of the electronic
thermal conductivity of graphene in the regime when \EE\ scattering is not
dominant.  (See, e.g., Ref.~\cite{PhysRevLett.115.056603} 
and Ref.~\cite{crossno:science}
for an in-depth analysis
on the hydrodynamic regime in which the \EE\ scattering is dominant.)

The electronic thermal resistivity of graphene $1/\kappa_e$ can be divided into two parts: 
the thermal resistivity arising from the \EP\ scattering, $1/\kappa_e^\EP$, and that from the impurity scattering, $1/\kappa_e^{\rm imp}$.
To calculate $1/\kappa_e^\EP$ we use the approach of Ref.~\cite{allen1978}, using a
variational solution of the Boltzmann transport equation:
\begin{equation}
\begin{split}
\frac{1}{\kappa_e^\EP} =& \frac{1}{L_{0} T} \frac{2\pi{}A\,d}{e^2\, N_{\rm{F}}\, v_{\rm{F}}^{2}} \int_{0}^{\infty} d\omega \frac{x}{{\rm sinh}^{2}x}
\\ 
&\Bigg[  \left( 1 - \frac{2x^2}{\pi^2} \right) \alpha_{\rm{tr}}^{2} F(\omega) 
+  \frac{6x^2}{\pi^2} \alpha^{2} F(\omega) \Bigg],
\end{split}
\end{equation}
where $L_0 = 2.44 \times 10^{-8}~~\rm{W\,\Omega/K^2}$ is the Lorenz number, $d=3.32$~\AA\ the inter-layer distance of graphite
(for comparison with bulk materials), $A$ the area of a unit cell of graphene,
$N_{\textrm{F}}$ the density of states per spin at the Fermi level, $v_{\textrm{F}}$
the Fermi velocity, and $x=\hbar\omega/2k_{\textrm{B}} T$.
The Eliashberg function $\alpha^{2} F(\omega)$ and the transport spectral function $\alpha_{\textrm{tr}}^{2} F(\omega)$ are given by
\begin{equation}
%\begin{split}
%\alpha^{2}F(\omega)=\frac{1}{N_{\textrm{F}}}\sum_{m,m',\nu} & \iint \frac{d%\mathbf{q}\, d\mathbf{k}}{A_{\textrm{BZ}}^{2}} \left| g_{m',m}^{\nu}(\mathbf{k,q}) %\right|^{2} \times \\
%& \delta\left(E_{\mathbf{k+q}}^{m'} - E_{\textrm{F}}\right)
%\delta\left(E_{\mathbf{k}}^{m} - E_{\textrm{F}}\right)
%\delta\left(\hbar\omega_{\mathbf{q}}^{\nu} - \hbar\omega\right),
%\end{split}
\begin{split}
\alpha^{2}F(\omega)=& \frac{1}{N_{\textrm{F}}}\sum_{m,m',\nu}  \iint \frac{d\mathbf{q}\, d\mathbf{k}}{A_{\textrm{BZ}}^{2}} \left| g_{m',m}^{\nu}(\mathbf{k,q}) \right|^{2} \times
\\
&\delta\left(E_{\mathbf{k+q}}^{m'} - E_{\textrm{F}}\right)
\delta\left(E_{\mathbf{k}}^{m} - E_{\textrm{F}}\right)
\delta\left(\hbar\omega_{\mathbf{q}}^{\nu} - \hbar\omega\right),
\end{split}
\end{equation}
and
\begin{equation}
\begin{split}
\alpha_{\textrm{tr}}^{2}F(\omega)=& \frac{1}{N_{\textrm{F}}}\sum_{m,m',\nu} \iint \frac{d\mathbf{q}\, d\mathbf{k}}{A_{\textrm{BZ}}^{2}} \left| g_{m',m}^{\nu}(\mathbf{k,q}) \right|^{2} \times \\
&\left(1 - \frac{v_{\mathbf{k+q}}^{m'} \cdot v_{\mathbf{k}}^{m} }{|v_{\mathbf{k}}^{m}|^{2}} \right) 
\delta\left(E_{\mathbf{k+q}}^{m'} - E_{\textrm{F}}\right) \times \\
&~\delta\left(E_{\mathbf{k}}^{m} - E_{\textrm{F}}\right)
\delta\left(\hbar\omega_{\mathbf{q}}^{\nu} - \hbar\omega\right),
\end{split}
\end{equation}
respectively, where $A_{\rm{BZ}}$ is the area of the Brillouin zone, $E_{\mathbf{k}}^{m}\,(E_{\mathbf{k+q}}^{m'})$ the electron energy with momentum $\mathbf{k\,({k+q)}}$
and band index $m\,(m')$, and $\omega_{\mathbf{q}}^{\nu}$ the frequency of a phonon
with momentum $\mathbf{q}$ and branch index $\nu$.
Also, the \EP\ matrix element $g_{m',m}^{\nu}(\mathbf{k}, \mathbf{q})$ is 
$\left< \mathbf{k+q}, m' \right| \delta{}V_{\rm{SCF}}(\mathbf{q},\nu) \left| \mathbf{k}, m \right>$,
where $\delta V_{\rm{SCF}}(\mathbf{q},\nu)$ is the derivative of the self-consistent potential $V_{\rm SCF}$
with respect to the ionic displacement by a phonon in branch $\nu$ with momentum $\mathbf{q}$.
As mentioned, all the \EP\ coupling matrix elements above are determined using an accurate
fit~\cite{park2014NL} of first-principles calculations incorporating \EE\
interactions at the level of the \textit{GW} approximation. A fit
is made necessary by the need to perform extremely dense Brillouin zone integrations, 
and both the first-principles physics and the quality of the fit have been validated in our study~\cite{park2014NL,Sohier2014PRB} of the electrical resistivity of doped graphene~\cite{efetov2010}.

\begin{figure}
%\centering
  \includegraphics[height=6cm]{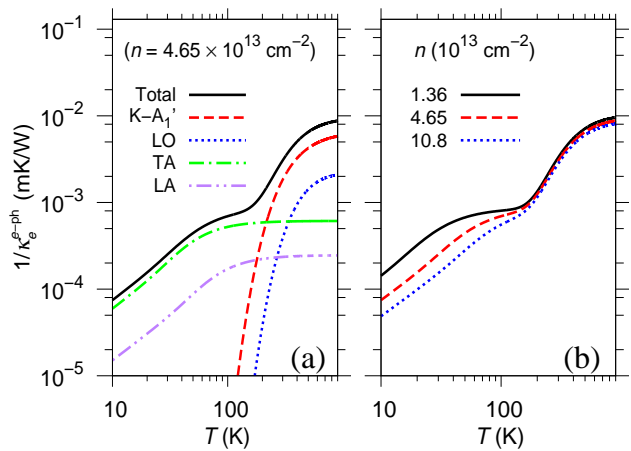}
  \caption{(a) Temperature dependence of the electronic thermal resistivity in doped graphene ($n=4.65 \times 10^{13}\,\rm{cm}^{-2}$) due to \EP\ interactions (solid black curve) as well as partial
contributions arising from different phonon modes.
(b) Temperature dependence of the electronic thermal resistivity of graphene
at different doping levels.
All the results are obtained with a model
that includes the effects of \EE\ interactions at
the level of {\it GW} many-body perturbation theory.}
  \label{fgr:fig_1}
\end{figure}

Figure~\ref{fgr:fig_1}(a) shows the first key result of the paper, with the calculated electronic thermal resistivity
$1/\kappa_e^\EP(T)$ as a function of temperature and doping, and its decomposition according to the contributions coming from each phonon branch.
As expected, at low temperatures,
the contribution from the acoustic phonon modes are dominant,
and proportional to $T$; the contribution from the transverse 
acoustic (TA) phonons is four times larger
than that from longitudinal acoustic (LA) phonons (the out-of-plane ZA phonons
have zero scattering by symmetry).
As the temperature increases, the partial contribution to
$1/\kappa_e^\EP(T)$ from the low-energy phonons increases
at a slower rate and eventually becomes independent of $T$
[Fig.~\ref{fgr:fig_1}(a)].
For the doping considered Fig.~\ref{fgr:fig_1}(a),
the crossover temperatures for LA and TA phonon branches are
$\sim30$~\% of the corresponding $T_{\rm{BG}}$'s, but these
crossover temperatures increase with
the doping density $n$ because $T_{\rm{BG}}\propto k_{\rm{F}}\propto\sqrt{n}$
[Fig.~\ref{fgr:fig_1}(b)].
(The crossover temperature of each phonon branch (either LA or TA) is obtained by extrapolating the low- and high-temperature behaviors to the intermediate-temperature regime and by taking the intersection of the two lines.)

Importantly, at $T > 200$~K, $1/\kappa_e^\EP(T)$ becomes dominated by
contributions from the longitudinal optical (LO) 
phonon mode (the contribution from the transverse optical phonon mode is negligible
as in the case of electrical resistivity~\cite{park2014NL}) and from the highest-energy
zone-boundary mode (K-A$_{1}'$) [Fig.~\ref{fgr:fig_1}(a)];
among these, the K-A$_{1}'$ phonon contribution is the largest.
In contrast to the case of acoustic-phonon scattering, the temperature dependence of the high-energy phonon scattering is hardly affected by the doping density, 
because the crossover temperature, $\sim0.3\,\hbar\omega^{\rm (LO,\,K-A_1')}/k_{\rm{B}}$, is independent of $n$.

\begin{figure}
\centering
  \includegraphics[height=11cm]{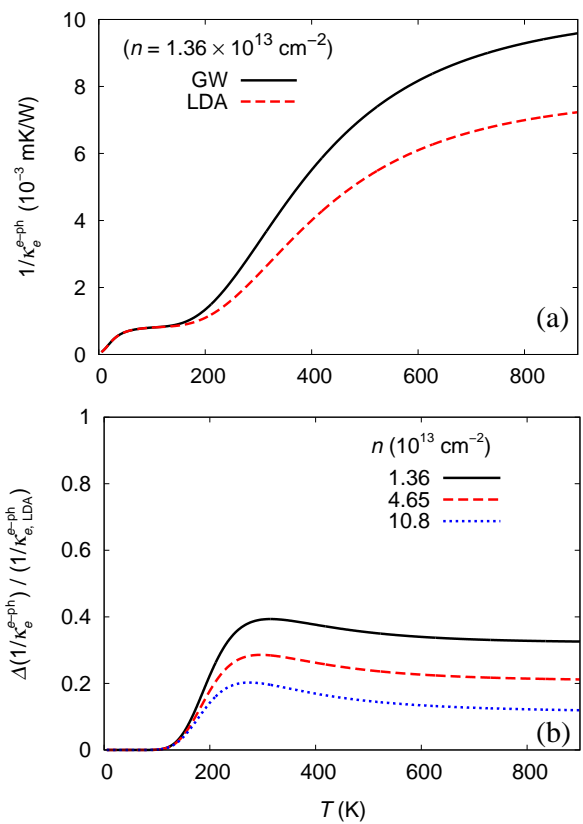}
  \caption{(a) Temperature dependence of the intrinsic electronic thermal resistivity of doped graphene ($n=1.36 \times 10^{13}\,\rm{cm}^{-2}$) calculated within density-functional theory (local-density approximation) or
with \textit{GW} many-body perturbation theory.  (b) The relative difference between
the two results, at different doping levels.}
  \label{fgr:fig_2}
\end{figure}

While these results include the effects of \EE\ correlations, we highlight in Fig.~\ref{fgr:fig_2}(a) the
role of these \textit{GW} corrections
to $1/\kappa_e^{\EP}$, as opposed to a pure density-functional description in the local-density approximation (LDA)~\cite{CeperleyAlder,PerdewZunger}.
The visible effect of correlations is the enhancement of the
contribution from the K-A$_{1}'$ phonon mode to $1/\kappa_e^{\EP}$,
via the renormalization of the \EP\ coupling matrix
elements~\cite{park2014NL}; on the other hand for phonon modes around the
zone center the enhancement of the Fermi velocity and that
of the \EP\ coupling matrix elements more or less cancel each other.
Therefore, the effect of the \textit{GW} correction can be clearly
seen at $T > 200$~K, where the contribution from K-A$_{1}'$ phonon
modes is the largest.
Figure~\ref{fgr:fig_2}(b) shows the doping-density dependence of the \textit{GW} correction to  $1/\kappa_e^{\EP}$: the correction is larger
at a lower charge density because the \EP\ coupling
is enhanced by weaker screening.

\begin{figure}
\centering
  \includegraphics[height=7cm]{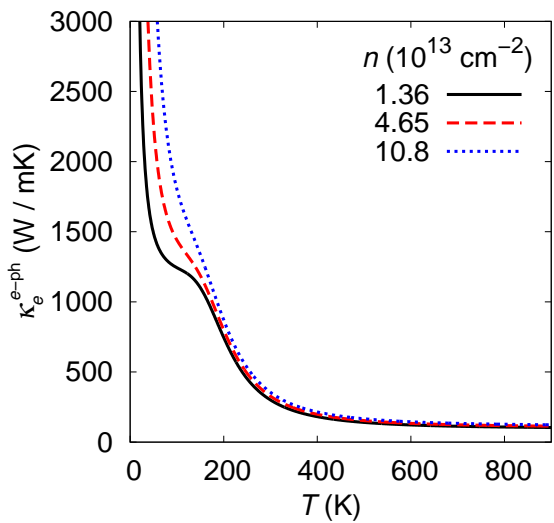}
  \caption{Temperature dependence of the intrinsic electronic thermal conductivity of doped graphene at different doping levels.}
  \label{fgr:fig_3}
\end{figure}

In order to facilitate comparison with experimental results, typically discussed in terms of 
conductivities, we show in Fig.~\ref{fgr:fig_3} the temperature dependence of 
$\kappa^{\EP}$, where it also become apparent that at low 
temperatures $\kappa^{\EP}(T)$ is proportional to $1/T$ (see also Fig.~\ref{fgr:fig_1})
and is higher at higher densities. Notably, at room temperature $\kappa^{\EP}$ is $\sim 
300$~W/mK, which is about 10~\% of the corresponding phonon
contribution and it is in absolute value corresponding to the total thermal conductivity of a typical bulk metal~\cite{balandin2008,chen_raman_2011,fugallo2014}.

While these results set the upper bound for $\kappa_e$, it is important also
to consider the contributions to the thermal resistivity of the electrons
arising from electron-impurity interactions,
$1/\kappa_e^{\rm imp}$. To do this, we use the measured electrical resistivity~\cite{efetov2010} as $T\to0$, which is considered
$1/\sigma^{\rm imp}$, and the Wiedemann-Franz law,
i.\,e.\,, $\kappa_e^{\rm imp} = L_0 T \sigma^{\rm imp}$.
Then, $\kappa_e$ is calculated using Matthiessen's rule:
$\kappa_e = \left(1/\kappa_e^{\rm imp} + 1/\kappa_e^{\EP}\right)^{-1}$.
\begin{figure}
\centering
  \includegraphics[height=7cm]{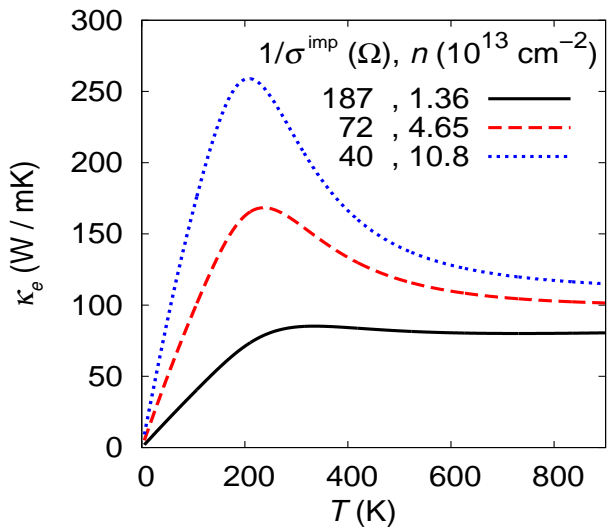}
  \caption{Temperature dependence of the electronic thermal conductivity
  of doped graphene at different doping levels, calculated including both the effects of impurities and of the \EP\ interactions. The effect of impurities is calculated using the experimental results of Ref.~\cite{efetov2010}.}
  \label{fgr:fig_4}
\end{figure}

Figure~\ref{fgr:fig_4} shows that in the low-temperature limit the impurity contribution to $\kappa_e$ is dominant and $\kappa_e(T) \propto T$, while
at high temperatures $\kappa_e(T)$ becomes independent of temperature,
which is typical also for bulk metals.
The calculated $\kappa_e$ at room temperature
is in the range of 80--200~W/mK and it is higher at a higher doping densities.

\begin{figure}
\centering
  \includegraphics[height=7cm]{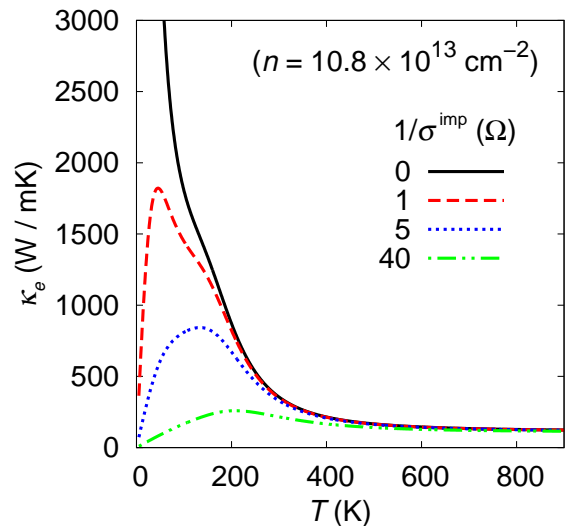}
  \caption{Temperature dependence of the electronic thermal conductivity in doped graphene
  ($n=10.8 \times 10^{13}\,$cm$^{-2}$) for different amount of impurities $ 1/\sigma^{\rm imp} $; the highest value $(40\,\rm{\Omega})$ is taken from the experimental results reported in Ref.~\cite{efetov2010}.}
  \label{fgr:fig_5}
\end{figure}

In Fig.~\ref{fgr:fig_5} we show the effects of the density of impurities on $\kappa_e$, considering
different residual electrical resistivities ($1/\sigma^{\rm imp}$) and the
validity of Wiedemann-Franz law for the electronic contributions
to the electrical and thermal conductivities limited by impurity
scatterings~\cite{ziman1961,PhysRevX.2.031006,apl.106.023121}, for $1/\sigma^{\rm imp}$
in a range between 0 and a maximum of
$40\,\rm{\Omega} $ that has been reported in Ref.~\cite{efetov2010} 
($1/\sigma^{\rm imp}$ largely depends on the sample condition, doping method,
and the substrate). We see that while
$\kappa_e$ is reduced significantly at low temperatures as a function of the
impurity density, most notably around and above room temperature
these effects are not very important.

\begin{figure}
\centering
  \caption{(Please contact the authors or visit Nano Letters web site for figures.) Temperature-dependent deviation from the Wiedemann-Franz law, plotted as $L(T) = \kappa_e(T)/\sigma(T)T$ for graphene at different doping levels
and at different $1/\sigma^{\rm imp}$ values.
Here, $ L_{0}$ is the Lorenz number.}
\label{fgr:fig_6}
\end{figure}

We now investigate the validity of Wiedemann-Franz law and discuss
the temperature dependent function
$L(T) = \kappa_e(T)/\sigma(T)T$ [obviously the validity of Wiedemann-Franz law is equivalent to stating that $L(T)/L_0=1$ independently of $T$].
It is known that the Wiedemann-Franz law holds if the elastic scattering is dominant (for example, when the temperature is low and impurity scattering is dominant,
the case that was used in the previous paragraph to estimate a typical amount of impurities in graphene from experimental data),
or for inelastic scattering in the high-temperature limit where all the phonons participate in electron scattering~\cite{ziman1961}.
At intermediate temperatures, due to the inelastic nature of the electron-phonon scattering, one could expect the largest deviations of $L(T)$ from $L_0$.
Our results are summarized in Fig.~\ref{fgr:fig_6}, where it can be seen that indeed, at 
low temperatures, $L(T)/L_{0}\approx 1$, given that $L(T)$ is determined mostly by the
electron-impurity scattering rather than \EP\ interactions;
small deviations are due to the acoustic-phonon contribution to $\kappa_e$ and $\sigma$.
At high-temperature $L(T)/L_{0}$ is again $\approx 1$, with $L(T)$ determined dominantly by \EP\ interactions, while the deviations from Wiedemann-Franz are the greatest around
room temperature, where $L(T)$ can be 20--50~\% lower than $L_0$, depending on the doping density.
Figure~\ref{fgr:fig_6} also shows that the violation of the Wiedemann-Franz law depends sensitively on $1/\sigma^{\rm imp}$ and in general is more severe in samples
with low $1/\sigma^{\rm imp}$, i.\,e.\,, in cleaner samples.

When the mean free path is longer than the characteristic length
of crystalline domain or sample, boundary scattering limits the thermal conductivity~\cite{LeeBroido2015NatComm,Cepellotti2015NatComm}.
The mean free path of electrons in graphene is given by 
$\Lambda_e = \sigma\,/e^2\, N_{\rm F}\, v_{\rm F}$.\cite{hwang2008} 
For the same $\sigma$ and charge density that we used in the calculation of $\kappa_e$ (Fig.~\ref{fgr:fig_4} and also Ref.~\cite{efetov2010}),
the mean free path of the electrons is $\sim$100~nm.
It is known that the lattice thermal conductivity of graphene is largely affected by the size of the sample~\cite{chen_raman_2011,fugallo2014, xu2014}.
The lattice thermal conductivity of a polycrystalline graphene with $1~\mu$m domains is a fraction of that of an infinitely large single-crystal graphene~\cite{fugallo2014}.
Since $\Lambda_e$ is an order of magnitude smaller than
1~$\mu$m, there are no such effects for the electronic thermal conductivity.
Therefore, the electronic thermal conductivity is relatively more important in graphene with small domains or in small graphene samples.

In summary, we calculate the electronic thermal conductivity of graphene
from first-principles
by fully taking into account the electron-phonon interactions
involving both low-energy, acoustic phonons and high-energy,
optical and zone-boundary phonons, and using both density-functional perturbation theory
at the LDA level, and \textit{GW} many-body perturbation theory.
We find that the electronic thermal conductivity in doped graphene without impurities
is $\sim300$~W/mK at room temperature, which is comparable to
the total thermal conductivity of a typical metal
and is $\sim$10~\% of that of graphene;
this value sets the upper bound of the electronic contribution to the thermal conductivity.
When we include the effect of impurities following the experimental results~\cite{efetov2010}, we find that the electronic thermal conductivity is
reduced and is of the order of 80--200~W/mK at room temperature, depending on the doping.
[In more resistive samples (see, e.\,g.\,,
Ref.~\cite{ghosh2008}), the electronic thermal conductivities could be lower.]
We also investigate the validity of the Wiedemann-Franz law in
the case of doped graphene, and find that 
$ L(T) = \kappa_e(T)/\sigma(T)T $ deviates from $L_0$ by
a 20--50\% at room temperature.
At low temperatures, the electronic thermal conductivity depends largely on the amount of impurities; however, above room temperature the impurity effects are small.
Because the mean free path of the electrons is short,
the electronic thermal conductivity is not significantly reduced even for sample sizes
as small as micron sizes, whereas the lattice thermal conductivity
is significantly reduced compared to that of an infinitely large
sample; hence, the electronic contribution to the thermal conductivity of graphene
samples as small as or smaller than
a few~$\mu$m is more important
than in the case of infinitely large, single-crystal samples.

%%%%%%%%%%%%%%%%%%%%%%%%%%%%%%%%%%%%%%%%%%%%%%%%%%%%%%%%%%%%%%%%%%%%%
%% The "Acknowledgement" section can be given in all manuscript
%% classes.  This should be given within the "acknowledgement"
%% environment, which will make the correct section or running title.
%%%%%%%%%%%%%%%%%%%%%%%%%%%%%%%%%%%%%%%%%%%%%%%%%%%%%%%%%%%%%%%%%%%%%
\begin{acknowledgments}
T.Y.K. and C.-H.P. acknowledge support from Korean NRF-2013R1A1A1076141 funded by MSIP and N.M. from Swiss NSF through grant
$200021\_143636$. Computer facilities were provided by
the 2014 Aspiring Researcher Program through Seoul National University.
\end{acknowledgments}

%%%%%%%%%%%%%%%%%%%%%%%%%%%%%%%%%%%%%%%%%%%%%%%%%%%%%%%%%%%%%%%%%%%%%
%% The appropriate \bibliography command should be placed here.
%% Notice that the class file automatically sets \bibliographystyle
%% and also names the section correctly.
%%%%%%%%%%%%%%%%%%%%%%%%%%%%%%%%%%%%%%%%%%%%%%%%%%%%%%%%%%%%%%%%%%%%%
%merlin.mbs apsrev4-1.bst 2010-07-25 4.21a (PWD, AO, DPC) hacked
%Control: key (0)
%Control: author (8) initials jnrlst
%Control: editor formatted (1) identically to author
%Control: production of article title (-1) disabled
%Control: page (0) single
%Control: year (1) truncated
%Control: production of eprint (0) enabled
%

\end{document}